# Evaluation of Frequent Itemset Mining Platforms using Apriori and FP-Growth Algorithm

*Ravi Ranjan, Aditi Sharma\**

*Department of Computer Science and Engineering, DIT University, Dehradun, India*

Abstract: With the overwhelming amount of complex and heterogeneous data pouring from any-where, any-time, and any-device, there is undeniably an era of Big Data. The emergence of the Big Data as a disruptive technology for next generation of intelligent systems, has brought many issues of how to extract and make use of the knowledge obtained from the data within short times, limited budget and under high rates of data generation. Companies are recognizing that big data can be used to make more accurate predictions, and can be used to enhance the business with the help of appropriate association rule mining algorithm. To help these organizations, with which software and algorithm is more appropriate for them depending on their dataset, we compared the most famous three MapReduce based software Hadoop, Spark, Flink on two widely used algorithms Apriori and Fp-Growth on different scales of dataset.

## 1. Introduction

According to www.worldwidewebsize.com, The Indexed Web contains at least 4.59 billion pages (Saturday, 29 July, 2017). With the massive proliferation in the velocity, volume and variety of information accessible online and the consequent need to develop viable paradigms which facilitate better techniques to access this information, there has been a strong resurgence of interest in Big data analysis research in recent years. Big data has enormous potential to improve the human condition, with emphasis on health and productivity

As stated on www.gartner.com, Big data is huge-volume, fast-velocity, and different variety information assets that demand innovative platform for enhanced insights and decision making. In other words, big data gets generated in multi-terabyte quantities, changes fast and comes in varieties of forms that is difficult to manage. IT organizations have started considering Big data initiative for managing their data in a better manner, visualizing this data, gaining insights of this data as and when required and finding new business opportunities to accelerate their business growth. Facebook, for example, knows how often you visit many websites (due to the pervasive Like on Facebook buttons) and wants to use that information to show you ads you are more likely to click on (Zahra and Nick, 2013).

To analyse such an unstructured data to get the information we want using computers is not easy. We need specialized techniques to perform such operations. Every major company has vast stores of information in increasingly complex databases. However, despite having more data than ever before, most data analytics still fail to provide actionable insights. Just as it is challenging for humans, transactional data makes association rule mining a challenging task for machines as well. Transactional datasets are typically extremely large, both in terms of the number of transactions as well as the number of items or features that are monitored (Yang and Liu, 2010).

Much work has been done to identify heuristic algorithms for reducing the number of itemsets to search (Zhou, Zhong, Chang, Huang and Feng, 2010). Perhaps the most-widely used approach for efficiently searching large databases for rules is known as Apriori. Introduced in 1994 by Rakesh Agrawal and Ramakrishnan Srikant, the Apriori algorithm has since become somewhat synonymous with association rule learning (Agrawal, Imielinski and Swami, 1993).

In this paper, we have Compared the two most famous association rule based algorithms MR-Apriori and Fp-Growth on three MapReduce platforms (Hadoop, Spark, Flink) on different scale of dataset to mine the frequent itemsets from these datasets. This comparison helps us to understand which algorithm works best on which software under which conditions.

## 2. Research Methodology

This section contains procedure of how the research has been conducted, Firstly the research questions has been defined and then the relevant studies has been searched on the digital libraries to find the existing algorithms and comparisons.

   *1. Research Questions*

This survey plans to discover the various techniques proposed to improve the Apriori algorithm in mining the frequent itemsets using association rules. For this, three inquiries were raised as takes after:
   RQ1: What are the different techniques that has been used for Parallel Apriori and Fp-Growth.
   RQ2: Which MapReduce frameworks has been used for implementing Apriori Algorithm and Fp-Growth.
   RQ3: What are the limitations of the existing techniques?
   RQ4: What is the performance of Apriori and FP growth against one another under different environment

   *2. Search Strategy*

This progression involves search terms, writing assets and search process. The search terms include Frequent Itemset Mining, Apriori Algorithm, Fp-Growth, Parallel Apriori, comparison, Hadoop, Flink, Spark, MapReduce and their synonyms. The search was conducted using AND, OR logic.

The writing assets for the pursuit of essential studies incorporate four electronic databases (IEEE Xplorer, ACM Digital Library, Science Direct, and Google Scholar). The search terms directed already were utilized to discover the papers in these four electronic databases.

## 3. Terminologies

### 3.1. Big Data

Big Data is considered as a game changer for many application domains. It is based on the premise that generating Big Data is cheap and feasible for enterprises, institutions and organizations widely.

With the fast-changing world, many techniques have been proposed to handle the big data. The most famous technologies used for big data analytics are Apache Hadoop, Apache Spark and Apache Flink as shown in Fig. 1. Big data is creating Big Impact on industries today. World's 50% of the data has already been moved to Hadoop – The Heart of Big Data (http://hadoop.apache.org/). It is predicted that by 2017, more than 75% of the world's data will be moved to Hadoop and this technology will be the most demanding in the market as it is now. Further enhancement of this technology has led to an evolution of Apache Spark – lightning fast and general-purpose computation engine for large-scale processing. It can process the data up to 100 times faster than MapReduce (Kitchenham and Charters, 2007). While Apache Flink is a streaming engine that can also do batches. So, at its core, Flink is more efficient in terms of low latency.

**Fig 1.** Big Data Technologies

### 3.2. Association Rules

Association rule analysis is used to search for interesting connections among a very large number of elements. Human beings are capable of such insight quite intuitively, but it often takes expert-level knowledge or a great deal of experience to do what a rule learning algorithm can do in minutes or even seconds (Amsterdamer, Grossman, Milo & Senellart, 2013). Additionally, some datasets are simply too large and complex for a human being to find the needle in the haystack.

4. The building blocks of a market basket analysis are the items that may appear in any given transaction. Groups of one or more items are surrounded by brackets to indicate that they form a set, or more specifically, an itemset that appears in the data with some regularity. Transactions are specified in terms of itemset, such as the following transaction that might be found in a typical grocery store:

### 3.3. MapReduce

Map-Reduce is a product system for effortlessly composing applications that process the vast amount of structured and unstructured data stored in the HDFS (Wang, Zhang & Chang, 2008). It processes the huge amount of data in parallel by dividing the job (submitted job) into a set of independent tasks. MapReduce programs are composed in a specific style influenced by useful programming builds, specifically figures of speech for processing data (http://www.philippe-fournier-viger.com/spmf/index.php?link=datasets.php). Here in map reduce we get input as a list and it changes over it into yield which is again a list. It is the heart of Hadoop.

Hadoop Map-Reduce is exceedingly versatile and can be utilized across numerous PCs. Numerous little machines can be utilized to process jobs that ordinarily couldn't be processed by a huge machine. Conceptually, Map-Reduce programs transform lists of input data elements into lists of output data elements (Li, Zeng, He & Shi, 2012). A Map-Reduce program will do this twice, using two different list processing idioms: Map and the second is Reduce

### 3.4. Apriori Algorithm

The name Apriori is derived from the fact that the algorithm utilizes a simple prior (that is, a priori) belief about the properties of frequent itemsets. The Apriori algorithm employs a simple a priori belief to reduce the association rule search space: all subsets of a frequent itemset must also be frequent. This heuristic is known as the Apriori property (Zaki, Parthasarathy, 1997). Using this astute observation, it is possible to dramatically limit the number of rules to be searched. For example, the set {motor oil, lipstick} can only be frequent if both {motor oil} and {lipstick} occur frequently as well (Honglie, Jun, Hongmei, 2011). Consequently, if either motor oil or lipstick is infrequent, any set containing these items can be excluded from the search. The Apriori algorithm uses statistical measures of an itemsets "interestingness" to locate association rules in much larger transaction databases.

### 3.5. Fp-Growth

FP-tree based frequent itemset mining technique, called FP-Growth, created by Han et al accomplishes high proficiency, in comparison to Apriori-like approach (Han, Pei and Yin, 2000). The FP-Growth technique embraces the divide-and-conquer system, utilizes just two full I/O scans of the database, and keeps away from iterative candidate generation. Frequent pattern mining consists of two steps:
- Building a compact data structure, FP Tree (frequent pattern tree), which stores more data in less space.
- Second is building of a FP-tree based pattern growth strategy to reveal every frequent pattern recursively.

The primary scan aggregates the support of every item and afterward chooses items that fulfil minimum support. This strategy produces frequent 1-itemsets and after that stores them in frequency descending order (Joy & Sherly, 2016). The second scan builds FP-tree. The FP-Tree is a compressed representation of the input.

### 3.6. Hadoop

Hadoop is an open source tool from the ASF – Apache Software Foundation. If certain functionality does not fulfil our requirement, we can change it according to our need. Hadoop provides an efficient framework for running jobs on multiple nodes of clusters. Cluster means a group of systems connected via LAN. Hadoop provides parallel processing of data as it works on multiple machines simultaneously. Hadoop works in master – slave fashion. There is a master node and there are n numbers of slave nodes where n can be 1000s. Master manages, maintains and monitors the slaves while slaves are the actual worker nodes (Dean & Ghemawat, 2004). Master should be deployed on good configuration hardware.

### 3.7. Spark

Apache Spark is a general-purpose & lightning fast cluster computing system. It provides high-level API. For example, Java, Scala, Python and R. Apache Spark is a tool for Running Spark Applications. Spark is 100 times faster than Bigdata Hadoop and 10 times faster than accessing data from disk (http://fimi.ua.ac.be/data/). Spark is written in Scala but provides rich APIs in Scala, Java, Python and R. It can be integrated with Hadoop and can process existing Hadoop HDFS data. Apache Spark was introduced in 2009 in the UC Berkeley R&D Lab, later it becomes AMP Lab. It was open sourced in 2010 under BSD license. In 2013 spark was donated to Apache Software Foundation where it became top-level Apache project in 2014 (http://spark.apache.org/).

### 3.8. Flink

Apache Flink is an open source platform which is a gushing information stream engine that stipulates communication, fault-tolerance, and data-distribution for distributed computations over data streams. Flink is one of the best project of Apache. Flink is a climbable data analytics framework that is fully well-matched to Hadoop (Wei, Ma, Zhang, Liu & Shen, 2014). Flink can easily execute both stream processing and batch processing.

Work on Flink started in 2009 at a technical university in Berlin under the stratosphere. It was incubated in Apache in April 2014 and became a top-level project in December 2014. Flink is a German word meaning swift/Agile. The logo of Flink is a squirrel, in harmony with Hadoop ecosystem.

The key vision for Apache Flink is to overcome and reduces the complexity that has been faced by other distributed data-driven engines. It is achieved by integrating query optimization, concepts from database systems and efficient parallel in-memory and out-of-core algorithms, with the MapReduce framework (Lan & Gita, 2013). As Apache Flink is mainly based on the streaming model, Apache Flink iterates data by using streaming architecture. The concept of an iterative algorithm is tightly bounded into Flink query optimizer. Apache Flink's pipelined architecture allows processing the streaming data faster with lower latency than micro-batch architectures (https://flink.apache.org/).

## 4. Literature Survey

Xiangyang presents an Apriori enhanced calculation of parallel association rules based on MapReduce in 2016. The strategy accomplishes its parallelization through the MapReduce structure, streamlines unique database to recreate the transaction record database and produces the frequent itemsets, and requests in rising the frequent item sets as per the support degree, at that point mines frequent item sets in the cluster.

Jian Guo et al presents CMR-Apriori calculation in 2013 which depends on the conventional Apriori calculation that consolidates Map/Reduce parallel execution, with Map/Reduce programming model and related encoding operation. Through twice Map/Reduce process, CMR-Apriori calculation enormously decreases the running time of the algorithm, tackling issues utilizing proficient and exact calculations.

Yihua Huang in 2014 proposed YAFIM (Yet Another Frequent Itemsets Mining), a parallel Apriori calculation in light of the Spark RDD structure an uncommonly composed in-memory parallel processing model to bolster iterative calculations and intuitive information mining. Experimental results demonstrate that, contrasted with the calculations implemented with MapReduce, YAFIM accomplished 18× speedup in normal for different benchmarks.

Sheng-Hui Liu in 2014, introduced an enhanced reformative Apriori calculation that uses the length of every transaction to decide the extent of the most extreme candidates itemset. By reducing the creation of low frequency itemset in Map function, memory depletion is enhanced, incredibly enhancing execution effectiveness.

Dachuan Huang proposes new upgrades to the MapReduce usage of FIM calculation by presenting a cache layer and a particular online analyzer in 2015. They assessed the adequacy and productivity of Smart Cache by means of broad trials on four open datasets. Smart Cache can lessen by and large 45.4%, and up to 97.0% of the aggregate execution time compared with the state-of-the-art solution.

Feng Gui in 2015 proposed DPBM, an appropriated framework construct pruning calculation situated in light of Spark, which manage FIM (frequent Itemsets mining). DPBM incredibly decrease the measure of candidate itemset by presenting a innovative pruning strategy for matrix-based frequent itemset mining algorithm, an enhanced Apriori calculation which just needs to check the input data once. What's more, every PC node lessens enormously the memory utilization by implementing DPBM under a most recent distributed-environment Spark, which is an exceptionally quick distributed computing. The exploratory outcomes demonstrated that DPBM have preferable execution time over MapReduce-based calculations on frequent itemset mining as far as speed and scalability is concerned.

Run-Ming Yu changed the conventional Apriori calculation by enhancing the execution productivity in 2014. Since the single-phase calculation just utilized only one MapReduce operation, it will produce unnecessary candidates itemset and result in deficient memory. He outlined and implemented a proficient algorithm: FPM (Frequent Patterns Mining) Algorithm solely based on MapReduce Framework (FAMR). He embraced Hadoop MapReduce as the investigation platform. The analysis comes about have demonstrated that FAMR is 16.2x faster in the run time contrasted with single-phase calculation.

Ning Li in 2012 implemented a parallel Apriori calculation in light of MapReduce, which is a structure for handling tremendous datasets on specific sorts of distributable issues utilizing countless number of computer nodes. The test displays that the proposed calculation can scale well and capably handle substantial huge datasets.

Xueyan Lin presented the MapReduce programming model of Hadoop platform and Apriori calculation of data mining, proposes the detailed steps of MR-Apriori calculation in 2014. Theoretical and experimental results indicated MR-Apriori calculation make a sharp increment in proficiency.

Zhuobo Rong in 2013 utilized the possibility of MapReduce parallel programming, the great Apriori and FP-Growth calculation are relocated to the MapReduce environment keeping in mind the end goal to effectively take care of the current issues of Apriori and FP-Growth calculation in the conventional usage techniques, and address the requirements of expansive data association rules mining.

## 5. Proposed Framework

Due to the increase in Web services and use of computers in most of the businesses, the amount of data available online and offline has changed drastically in terms of volume as it has become a global source of useful information. Analysing such an amount of data manually is impossible, so the researchers has proposed various algorithms to analyse and for mining the patterns available in data to predict the current trends in the society going on. Some platforms have also been developed to ease this process for organizations. But each organization has different type of data in terms of size, rate of increase of data, number of attributes, structured or unstructured data. Not every algorithm or platform is suitable in all the circumstances.

To understand the applicability of MapReduce Frameworks, Apriori and FP-Growth were implemented on three different datasets of different sizes on Apache Hadoop, Apache Spark and Apache Flink. The results of these two algorithms are compared on all three platforms on three different dataset conditions to recognize the conditions suitable for each algorithm in different situations. It just doesn't add value only to the advertising industry, the research community has long looked for systems to successfully disperse new scientific discoveries and technological breakthroughs so, as to propel our aggregate information and raise our progress.

The system firstly retrieves the data from online stores. The data is converted into item sets, each item set represents items brought together. Both the association rule mining algorithms are run on these three datasets separately on each of the MapReduce platforms. The time taken to generate the frequent items are recorded for each of the six implementations during the three trails for different data sets. The average of the three iterations are averaged to conclude the algorithm that performs good in all the conditions. Figure shows the overview of the system proposed in this research.

*Data Set*: The data set used in this research is collected from SPMF- An open source data mining library. It is a repository which specialized in pattern mining. We have taken three data sets from this data mining library of different size. The first data set is named as 'Food Mart', it is a dataset of customer transactions from a retail store, it contains 4141 entries. Entries are regarding 1554 different items. Second dataset is termed as 'Online Retail', it is transformed from the Online retail dataset, it contains 541909 transaction entries of 2603 items.

*MapReduce:* To evaluate the performance of the different MapReduce frameworks 'time' is taken as the evaluation metric to fetch the frequent itemsets.

## 6. Result and Analysis

For Food Mart dataset with around 4000 transaction and minimum support of 0.1%, Hadoop takes approximately 26 seconds whereas Spark and Flink takes 20 and 11 seconds, respectively.

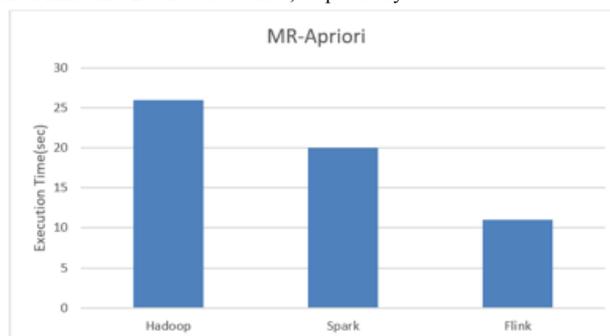

**Fig 3. Food Mart with Minimum Support 0.1%**

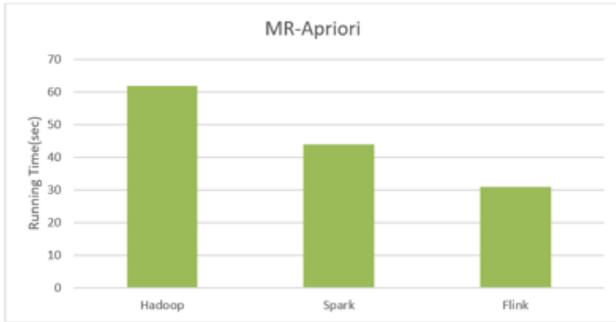

**Fig 4. T1014D100K with Minimum Support 0.3%**

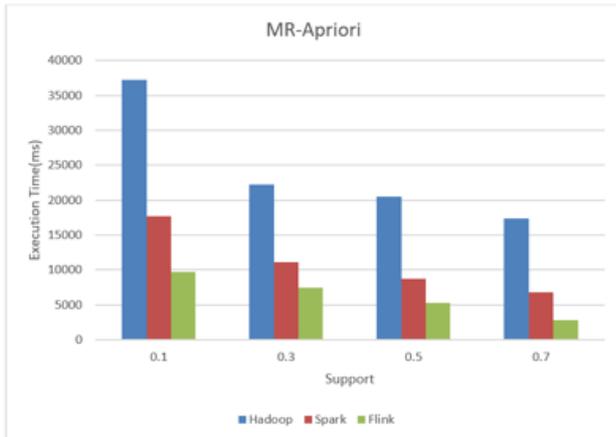

**Fig 5. Online Retail with Minimum Support 0.5%**

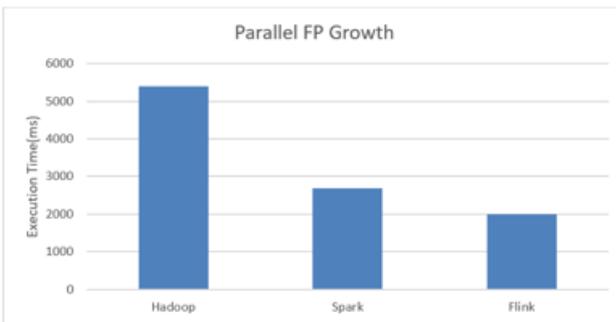

**Fig 6. Food Mart with Minimum Support 0.3%**

For T1014D100K dataset with around 100000 transaction and minimum support of 0.3%, Hadoop takes approximately 61 seconds whereas Spark and Flink takes 44 and 31 seconds, respectively.

Figure 5 shows the scalability of MR Apriori algorithm on three different platforms for the Online Retail dataset with around 500000 transaction and minimum support of 0.5%. Figure 21 compares the performance of Parallel FP Growth algorithm on three different platforms. Figure 7 analyses the scalability of PFP on Spark and Flink on dataset T1014D100K with minimum support of 0.5%.

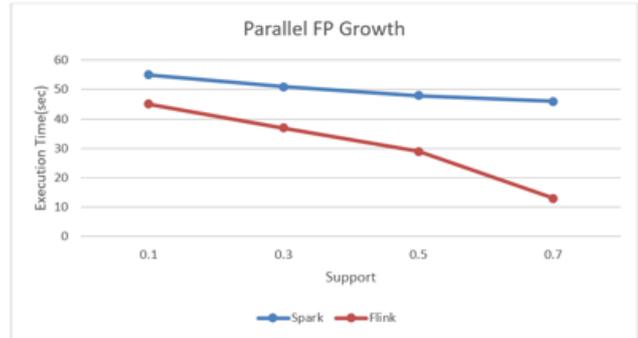

**Fig 7. T1014D100K with Minimum Support 0.5%**

## 7. Conclusion

The study in this work presents Flink based MR Apriori and Parallel FP Growth which is applied to mine frequent patterns from extensive datasets. It utilizes essential Apriori requirement that an itemset must be frequent if only all its non-empty subset is frequent. It is executed on Apache Flink, Apache Spark and Apache Hadoop which gives parallel and distributed processing condition. Flink is most appropriate for Apriori since Apache Flink have local support for iterative calculation and Apriori is based upon iterative calculation. Flink's pipelined design enable us to begin another Apriori iteration when few results of previous iteration are available. Delta cycle usefulness of Flink makes Apriori exceptionally parallel and powerful calculation for colossal datasets. In Summary, we have presented an execution of Apriori and FP Growth on Hadoop, Spark and Flink and tried to compare with various datasets. We also demonstrated that Flink based Apriori is equipped for dealing with extensive transactional-based datasets effortlessly.

Since Flink is the one of the most recent advancement in the field of Big data, not much work has been conducted to see how it performs with other distributed platforms. Also, not even a single algorithm in the field of association rule mining is introduced till date. Our work can be extended to cover large computer clusters dataset with more than one Tera bytes. Additionally, we can apply this parallel version of Apriori and FP Growth to various application domain such as weather data, internet traffic, medical information etc. We can also use these algorithms to generate different and interesting association rule faster and effectively.